\advance\topmargin by 2.0in
\documentclass[aps,prl,twocolumn,showpacs]{revtex4}
\usepackage{graphicx}
\usepackage{epsfig}

\begin{document}

\title{Exact exponents of edge singularities in dynamic correlation functions of
  1D Bose gas}

\author{Adilet Imambekov and Leonid I. Glazman }
\affiliation{Department of Physics, Yale University, New Haven,
Connecticut, USA, 06520}

\date{\today}

\begin{abstract}
  The spectral function and dynamic structure factor of bosons
  interacting by contact repulsion and confined to one dimension
  exhibit power-law singularities along the dispersion curves of the
  collective modes. We find the corresponding exponents exactly, by
  relating them to the known Bethe ansatz solution of the Lieb-Liniger
  model. 
  Remarkably, the Luttinger liquid theory
  predictions for the exponents fail even at low energies, once the
  immediate vicinities of the edges are considered.

\end{abstract}

 \maketitle

\def\be{\begin{equation}}
\def\ee{\end{equation}}
\def\bea{\begin{eqnarray}}
\def\eea{\end{eqnarray}}

In a typical experiment with cold atoms, the atomic ensemble is
kept isolated from environment over certain time. This decoupling
from the environmental degrees of freedom is effective enough to
make the interaction between the atoms to be the leading cause of
the ensemble's evolution in time. The isolation of ${}^{87}$Rb
atoms in elongated traps recently showed~\cite{dweiss} the
peculiarity of the dynamics of bosons confined to one dimension
(1D),  showing almost no relaxation, and consistent with the
notion of the integrability.


Integrability allows one to find the spectrum of excitations of
the quantum system by means of a thermodynamic Bethe ansatz (TBA),
as it was done for the case of pointlike interaction between
bosons by Lieb~\cite{LL}. A more recent progress based on
algebraic Bethe ansatz~\cite{KBI} ideas helped to develop
sophisticated numerical methods for studying the dynamic
responses of 1D Bose gas~\cite{CauxDSF, CauxG}. However, in spite
of more than 40 years of research, the analytic calculation of
correlation functions from Bethe ansatz wave functions still
remains a challenge.
Some progress in understanding of dynamical correlations was
achieved recently along a different path, which uses the
effective Hamiltonian description and the physical analogy with
the Fermi edge
singularity~\cite{Pustilnik2006Fermions,Khodas2006Fermions,Khodas2007Bosons,Khodas2007Soliton}.
The analogy helped to show that the dynamic structure factor (DSF)
and spectral function may exhibit power-law singularities along
the dispersion curves of the collective modes.
Very recently Pereira {\it et al.}~\cite{Affleck2007} put forward
an idea to combine the effective Hamiltonian method developed in
Refs.~\cite{Pustilnik2006Fermions,Khodas2006Fermions,Khodas2007Bosons}
with the TBA method to study the edge singularities of XXZ spin
chain. In their approach, Pereira {\it et al.} used the ideas
based on conformal field theory to relate the parameters of the
effective Hamiltonian to the finite-size corrections to the
energy.

Here we find an alternative way to directly relate the parameters
of the effective Hamiltonian to the scattering phase shifts found
in the Bethe ansatz solution of the Lieb-Liniger model. We extend
the formalism of
Refs.~\cite{Pustilnik2006Fermions,Khodas2006Fermions,Khodas2007Bosons}
to evaluate the exponents not only of the DSF, but of the spectral
function as well. In the low energy limit we analytically evaluate
exact exponents as a function of Luttinger parameter $K$ only.
Remarkably, the Luttinger liquid (LL) theory
predictions~\cite{EL,Haldane,Caza04,Giamarchibook} for the
exponents fail even at low energies, once the immediate vicinities
of the edges are considered. Our results at low energies help
understanding the behavior of 1D systems beyond the linear
spectrum approximation of the LL theory.
They can be used as a benchmark for numerical methods which
attempt to evaluate many-body dynamics of continuous 1D models,
e.g., using t-DMRG algorithms~\cite{DMRG}.


DSF describes the probability to excite the ground state with
momentum and energy transfer $(k,\omega),$ and can be measured
using Bragg spectroscopy~\cite{MITbraggscattering}. Spectral
function  describes the tunneling probability for a particle (or a
hole) with momentum and energy $(k,\omega)$ respectively. It can
be measured using stimulated Raman transition combined with
additional spin-flips~\cite{Duan, Carusotto} .

In what follows we will be interested in the zero-temperature DSF
\bea S (k,\omega) = \int\!dx\,dt\,e^{i (\omega t-kx)}\, \bigl
\langle \rho (x, t) \rho (0, 0) \bigr \rangle, \label{DSF} \eea
 and spectral function
$ A(k,\omega)= -\frac1{\pi}{\rm Im }G(k, \omega)\, {\rm
sign}\omega,
$
where Green's function $G(k,\omega)$ is defined by \cite{AGD} \bea
G(k, \omega)= -i \int\int dx dt e^{i (\omega t -kx )} \bigl
\langle T\left(\Psi(x,t)\Psi^{\dagger}(0,0)\right) \bigr \rangle .
\eea Here $\Psi(x,t)$ and $\rho (x, t)$ are boson annihilation and
density operators, and $T$ denotes  time ordering. 
Energy $\omega$ is measured from chemical potential, so
$A(k,\omega)$ for $\omega>0 \;(\omega<0)$ describes the response
of the system to an addition of an extra particle (hole). Both
$A(k,\omega)$ and $S(k,\omega)$ do not change under transformation
$k\rightarrow -k,$ and we will consider them only for $k>0.$ Also
$S(k,\omega)=0$ for $\omega<0,$ so we
consider DSF
only for $\omega>0.$

\begin{figure}
\includegraphics[width=8cm]{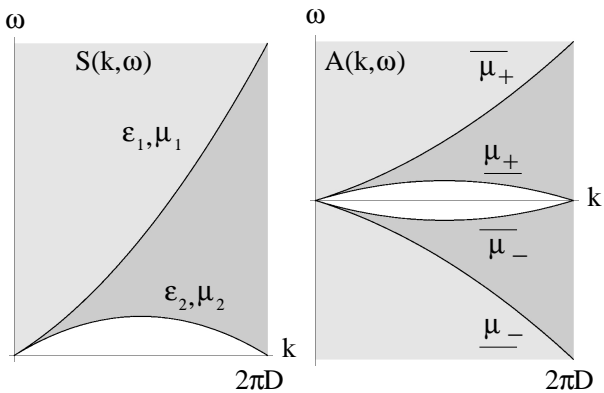}
\caption{\label{Fig1}(a) Dynamic structure factor (DSF)
$S(k,\omega)$ and (b) spectral function $A(k,\omega).$ Shaded
areas indicate the regions  where they are nonvanishing. Lieb's
particle mode $\varepsilon_1(k)$ and hole excitation mode
$\varepsilon_2(k)$ are indicated. For spectral function
$A(k,\omega)$ region with $\omega>0\; (\omega<0)$ corresponds to
the particle (hole) part of the spectrum. Notations of $\mu$
indicate which exponents of Eqs. (\ref{mu12})-(\ref{mudown})
should be used in Eq. (\ref{mudef}). }
\end{figure}

The exactly solvable Lieb-Liniger model \cite{LL, KBI} is defined
by \bea H_{LL}=-\sum_{j=1}^{N} \frac{\partial^2}{\partial z_j^2} +
2c\sum_{1\leq j<k\leq N} \delta(z_j - z_k) - h N. \label{LL}
 \eea
Here $N$ is the total number of particles, $h$ is the chemical
potential, $c>0$ is the interaction strength. Hereinafter we set
$\hbar=1$ and the mass equals $1/2$ for brevity.
The ground state is fully characterized by the dimensionless
parameter $\gamma = c/D,$ where $D=N/L$ is the density. The regime
of weak interactions corresponds to $\gamma\ll 1,$ while the
strong interactions (Tonks-Girardeau limit) correspond to $\gamma
\gg 1.$ An important parameter appearing in the effective
hydrodynamic description \cite{EL,Haldane,Caza04} of the model
(\ref{LL}) is the LL parameter $K=v_f/v,$ where $v$ is the sound
velocity and $v_f=2\pi D$ is the Fermi velocity of noninteracting
Fermi gas of density $D.$ Parameter $K>1$ is uniquely defined by
$\gamma,$ with $K\approx \pi\gamma^{-1/2}$ for $\gamma\ll 1$ and
$K\approx 1+4/\gamma$ for $\gamma\gg 1$~\cite{Caza04,LL}.
Energies of excitations can be also obtained, and give rise to
Lieb's particle $(\varepsilon_1)$ and hole $(\varepsilon_2)$
excitations, as shown in Fig.\ref{Fig1}. The minimal excitation
energy (measured from the ground state) of a state with total
momentum $k$ equals $\varepsilon_2(k),$ and both $A(k,\omega)$
and $S(k,\omega)$ vanish identically for
$|\omega|<\varepsilon_2(k).$

 At
zero temperature, $S(k,\omega)$ and $A(k,\omega)$ have power-law
behavior
\cite{Khodas2006Fermions,Khodas2007Bosons,Pustilnik2006Fermions}
near $\pm \varepsilon_{1(2)}(k),$ \be
 S(k,\omega), A(k,\omega) \sim {\rm const}+ \left|\frac{1}{\omega \pm
 \varepsilon_{1(2)}(k)}\right|^{\mu}, \label{mudef}
\ee with notations of $\mu$ shown in Fig. \ref{Fig1}.
We provide exact results for exponents at $\pm
\varepsilon_{1}(k)$ with $k>0,$ and for exponents at $\pm
\varepsilon_{2}(k)$ with $0<k<2\pi D,$   see
Eqs.~(\ref{mu12})~-~(\ref{nonluttexp6}).

Let us briefly review the solution of the Lieb-Liniger model to
set the notations, and we will mostly follow the conventions of
Ref. \cite{KBI}. Ground state quasimomenta $\nu_j, 1\leq j \leq
N,$ are given by the solutions of Bethe equations \bea L \nu_j +
\sum_{k=1}^N \theta (\nu_j-\nu_k)=2\pi n_j, \label{Betheeqs} \eea
where $\theta(x)=2 \arctan{\frac{x}{c}}$ is the two-particle
phase shift and quantum numbers are $n_j=j-1 -(N-1)/2.$ In the
thermodynamic limit, this system gives rise to the integral
equation \bea 2\pi \rho(\nu)- \int_{-q}^{q} K(\nu, \mu) \rho(\mu)
d \mu= 1. \eea
 Here $\rho(\nu)=\lim 1/(L(\nu_{k+1}- \nu_k))$ is the
density of roots, $K(\nu, \mu)=\frac{2c}{c^2+(\nu - \mu)^2},$ and
$\pm q$ is the highest (lowest) filled quasimomentum; $q$ is
defined as a function of density by the normalization condition $
D=\int_{-q}^{q}\rho( \nu) d\nu.
$ Particlelike excitations  with $k>0$ can be constructed by
adding an extra quasimomentum $\lambda>q,$ while holelike
excitations are obtained by removing a quasimomentum
$|\lambda|<q$ (particlelike states with $k<0$ correspond to
$\lambda<-q$). Since all quasimomenta $\nu_j$ are coupled to each
other by Eqs. (\ref{Betheeqs}), this will shift all of them. A
convenient way to take this change into account is to introduce a
shift function \bea F_B(\nu|\lambda)= \pm (\nu_j-\tilde
\nu_j)/(\nu_{j+1}-\nu_{j}), \label{FBdef} \eea where $\tilde
\nu_j$ are new solutions and upper (lower) sign corresponds to
extra particle (hole). In the thermodynamic limit
$F_B(\nu|\lambda)$ satisfies an integral equation \cite{fnote}
 \bea F_B(\nu|\lambda) -
\frac{1}{2\pi}\int_{-q}^{q} K(\nu, \mu) F_B(\mu|\lambda)d \mu=
\frac{\pi + \theta(\nu - \lambda)}{2\pi} \label{FBeq}.\eea Shift
function $F_B(\nu|\lambda)$ can be used~\cite{KBI} to calculate
exact energies $\varepsilon_{1(2)}$ as a function of momentum
$k(\lambda).$ The latter can be also written \cite{KBI} as \bea
k(\lambda)=\pm \left( \lambda - \pi D+  \int_{-q}^{q}
\theta(\lambda - \nu) \rho(\nu) d \nu\right).
\eea Here upper (lower) sign corresponds to particle (hole)
excitation with $\lambda>q \;(|\lambda|<q),$ and $k(q)=0, \;
k(-q+0)=2\pi D.$


 As will be shown below, $ F_B(\pm q| \lambda)$ play a crucial
role in the calculation of
 the edge singularities, so we will investigate it in more detail. One can analytically derive the
limiting behavior \bea
 F_B(\pm q| q)
=1/2 \pm \left(1/2 -1/(2\sqrt{K})\right),\label{F1exact} \eea and
$F_B(\pm q| -q)=\sqrt{K}-F_B(\mp q|q);$ in addition, \bea F_B(\pm
q| \lambda)\approx c\sqrt{K}/(\pi k)\ll1  \;\; \mbox{for}\;\; q,c
\ll\lambda\approx k. \label{F0exact} \eea
Eq. (\ref{F1exact}) can be derived \cite{fnote} from  Ref.
\cite{KorepinSlavnov}. Eq. (\ref{F0exact}) follows from expansion
of right hand side of Eq. (\ref{FBeq}) combined with  $\rho(\pm q)
=\sqrt{K}/2\pi,$ see e.g. Eqs. (I.9.20-I.9.22) of Ref. \cite{KBI}.

\begin{figure}
\includegraphics[width=8cm]{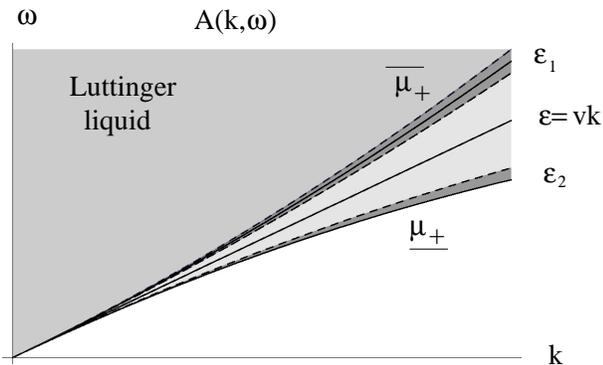}
 \caption{\label{Fig2} Spectral function
$A(k,\omega)$ for $\omega>0$ near $k=0.$ Exact exponents
$\overline{\mu_{+}}=1/\sqrt{K}-1/(2K),
\underline{\mu_{+}}=-1/\sqrt{K}-1/(2K)$ are valid only in the
immediate vicinities of $\varepsilon_1$ and $\varepsilon_2$
marked by dashed lines. Luttinger liquid behavior given by Eqs.
(\ref{Luttexp}) is valid only sufficiently far away from
$\varepsilon=vk,$ for $\omega- v k\gg \varepsilon_1
-\varepsilon_2.$ }
\end{figure}

We can calculate the exponents using the method of Refs.
\cite{Pustilnik2006Fermions,Khodas2006Fermions,Khodas2007Bosons,Affleck2007}.
For example, exponents at $\varepsilon_1(k)>0$ are evaluated using
effective Hamiltonian $H=H_0+H_d + H_{int},$ where \bea H_0=
\frac{v}{2\pi}\int dx\;\left( K (\nabla \theta)^2 +
\frac1{K}(\nabla \phi)^2\right ), \label{hbosonic}\\
 H_d=\int dx\; d^{\dagger} (x) (\varepsilon_1(k)-i \frac{\partial
 \varepsilon_1(k)}{\partial k } \frac{\partial}{\partial x})d(x), \\
 H_{int}=\int dx \left( V_{R} \nabla \frac{\theta-\phi}{2\pi}-V_{L}\nabla\frac{ \theta+\phi}{2\pi}\right) d^{\dagger} (x)
 d(x).\label{hint}
 \eea
Here $d^{\dagger}$ creates an extra particle with momentum near
$k,$ and we use the notations of Ref. \cite{Giamarchibook}, such
that $\Psi_B(x,t) \sim e^{i\theta(x,t)}, [\phi(x),\nabla
\theta(x')]=i \pi \delta(x -x').$ The effective Hamiltonian, Eqs.
(\ref{hbosonic})-(\ref{hint}), discriminates states created by
$d^\dagger$ from low-lying states of the rest of the system which
are described as LL. This separation is possible only in the
investigation of response functions very close to the Lieb's
modes, such as line $\varepsilon_1(k)$ in Fig. \ref{Fig2}, and
only due to the finite curvature of the underlying boson
spectrum. Under these circumstances, excitations close to the
edge can be distinguished from the excitations of the LL.
Response functions evaluated from Eqs.
(\ref{hbosonic})-(\ref{hint}) are valid in a narrow region of
width vanishing as $\sim k^2$ at $k\to 0;$ see discussion after
Eq. (\ref{Luttexp}).

Singular parts of
DSF and spectral function are 
\bea S(k,\omega) \sim \int
dx dt e^{i\omega t} \langle d \Psi(x,t) \Psi d^{\dagger}(0,0)
\rangle_{H_0+H_d+H_{int}},\nonumber \\ A(k,\omega)\sim \int dx dt
e^{i\omega t} \langle d(x,t) d^{\dagger}(0,0)
\rangle_{H_0+H_d+H_{int}}.\nonumber \eea

 The crucial step in our approach is the identification of
the unitary transformation which removes an interaction term
$H_{int}$ from the Bethe ansatz solution. This can be done by
noticing that $H_0$ becomes a sum of noninteracting modes after
transformation $\phi = \sqrt{K} \tilde \phi, \theta = \tilde
\theta/\sqrt{K}.$ If one refermionizes fields $\tilde \phi,
\tilde \theta,$ one obtains a noninteracting Luttinger model with
two branches. Fermionic excitations of this model with momenta
$k_j>0$ correspond to low-energy particlelike excitations of the
Bethe ansatz with quasimomenta $\nu_j>q,$ and $k_j \propto \nu_j
-q.$ The state where a $d$ particle is present corresponds to a
state with occupied quasimomentum $\lambda.$ For noninteracting
fermions, phase shift on the particle $d$ can be written as
$\delta(k)=2\pi (k_{j}- \tilde k_j)/(k_{j+1}-k_j),$ where $\tilde
k_j$ is the momentum of the new eigenstate in the presence of
particle $d.$ The same quantity in the  Bethe ansatz solution can
be calculated using Eq. (\ref{FBdef}), and by noticing that the
shift of quasimomenta by $\nu_{j+1}-\nu_j$ corresponds to a phase
shift $2\pi.$ Then interaction term $H_{int}$ can be removed
\cite{Schotte,Balents2000} by unitary transformation
$U^{\dagger}(H_0+H_d+H_{int})U,$ where \bea U^{\dagger}=e^{i\int
dx \left(\frac{\delta_+}{2\pi}(\tilde \theta(x)-\tilde \phi(x))
-\frac{\delta_-}{2\pi}(\tilde \phi(x)+\tilde
\theta(x))\right)d^{\dagger} (x)
 d(x) },\nonumber \\ \delta_{\pm}=2\pi F_B(\pm q,\lambda).\eea

A standard calculation
\cite{Giamarchibook,GNT,Schotte,Balents2000} then leads to \bea
 \mu_{1,2} =1 -
\frac12\left(\frac{1}{\sqrt{K}}+\frac{\delta_{+}-\delta_-}{2\pi}\right)^2-\frac12\left(\frac{\delta_{+}+\delta_-}{2\pi}\right)^2
\label{mu12},\\
 \overline{ \mu_{\pm}} =1
-
\frac12\left(\frac{\delta_{+}-\delta_-}{2\pi}\right)^2-\frac12\left(\frac{\delta_{+}+\delta_-}{2\pi}\right)^2
\label{muup}. \eea


Exponents $\underline{\mu_{+}}$ and $\underline{\mu_{-}}$ are more
complicated. For example, $\underline{\mu_{+}}$ corresponds to a
state with one additional particle, total momentum $k,$ and the
smallest possible energy $\varepsilon_2(k).$ Such state is given
by two extra particles at the right quasifermi surface and a hole.
To evaluate $\underline{\mu_{+}},$ one needs to calculate the
correlator $\langle d^{\dagger}\Psi\Psi (x,t)
\Psi^{\dagger}\Psi^{\dagger}d(0,0)\rangle,$ where $d$ creates a
hole.
One
obtains \bea
 \underline{\mu_{\pm}} =1
-
\frac12\left( \frac{2}{\sqrt{K}}+
\frac{\delta_{+}-\delta_-}{2\pi}\right)^2-\frac12\left(\frac{\delta_{+}+\delta_-}{2\pi}\right)^2
\label{mudown}.
 \eea
In Eqs.(\ref{mu12})-(\ref{mudown}) one should choose $\lambda>q $
for $\mu_1, \overline{ \mu_{+}}, \underline{\mu_{-}}$, and
$|\lambda|<q$ for $\mu_2, \underline{ \mu_{+}},
\overline{\mu_{-}}.$


Using Eqs. (\ref{F1exact})-(\ref{F0exact}), one can derive
analytically the behavior of exponents near $k=0, 2\pi D$ and at
$k\rightarrow \infty.$ DSF exponents are given by
 \bea \mu_{1,2}(0)=0,\\\
\mu_2(2\pi D-0)=2\sqrt{K}-2K<0,\label{nonluttexp1}\\
  \mu_1(\infty)=1-1/(2K)>0. \label{Khodasexp} \eea
Exponents for the spectral function at low energies equal
 \bea
\overline{\mu_{\pm}}(0)=1/\sqrt{K}-1/(2K)>0,\label{nonluttexp2}\\
\underline{\mu_{\pm}}(0)=-1/\sqrt{K}-1/(2K)<0,\label{nonluttexp3}\\
\overline{\mu_{-}} (2\pi D -0)= -2K + 2\sqrt{K} +
1/\sqrt{K}-1/(2K),\\ \underline{\mu_{+}}(2\pi D -0)= -2K +
2\sqrt{K} - 1/\sqrt{K}-1/(2K),\label{nonluttexp5}\eea and for
large momenta one has \bea \underline{\mu_{-}}(\infty)=1-2/K.
\label{nonluttexp6} \eea For $k\rightarrow \infty$ Eq.
(\ref{muup}) gives $\overline{\mu_{+}}\rightarrow 1,$ which is a
certain way to parameterize an expected result $A(k,\omega)\sim
\delta(\omega-k^2),$ since $\delta(\omega-k^2)\sim
\lim_{\overline{\mu_{+}}\rightarrow 1}\int dt e^{i (\omega-k^2)
t}/t^{1-\overline{\mu_{+}}}.$

 Numerical solutions show that all
exponents are monotonic functions of $k,$ lying between limiting
values cited above. Exponents $\overline{\mu_{-}}$ and
$\underline{\mu_{-}}$ can change sign as a function of $k.$
Exponent $\overline{\mu_{-}}$ changes its sign from positive at
$k=0$ to negative at $k=2\pi D$ for sufficiently weak
interactions, at $K>K_c=1/4 + \sqrt{5}/4 + 1/2\sqrt{1/2 +
\sqrt{5}/2} \approx 1.445,$ and $\underline{\mu_{-}}$ can also
change its sign from negative at $k=0$ to positive at $k=\infty$
for $K>2.$ Our results are in full agreement with previously
published limiting cases
\cite{Khodas2007Bosons,Khodas2007Soliton}, and agree
qualitatively with the numerical calculations of DSF
\cite{CauxDSF} and spectral function \cite{CauxG}.

 One of the main achievements of the LL approximation \cite{EL,Haldane,Caza04,Giamarchibook, GNT} is the
calculation of exponents in dynamic correlation functions in the
limit $k\rightarrow 2 \pi D n,$ where $n$ is any integer. We note
however that exact exponents
(\ref{nonluttexp1})-(\ref{nonluttexp5}) in the immediate vicinity
of the edges are different. They  show markedly non-LL behavior,
e.g. they can depend on $\sqrt{K},$ while LL exponents depend only
on $K$ and $1/K.$ To illustrate the failure of LL exponents in the
immediate vicinity of the edge we consider $A(k,\omega)$ for
$\omega>0$ near $k=0$ in more detail; see Fig. \ref{Fig2}. LL
theory assumes linear spectrum $\varepsilon (k)\approx v k$ and
predicts \cite{Caza04,Giamarchibook} at $k\ll v$
 \bea
 A(k, \omega) \sim \left\{ \begin{array}{c}
  0, \; \mbox{if } \; \omega<v k ,\\
  (\omega - v k )^{1/(4K)-1}\;\; \mbox{if } \; 0<\omega-v k\ll \omega , \\
  \omega ^{1/(2K)-2}\; \mbox{if } \; v k \ll
  \omega .\end{array} \right. \label{Luttexp}
 \eea
Difference between the second and  third lines here arises from
the fact that only one branch contributes to the exponent for
$0<\omega-v k\ll \omega,$ while both left- and right-movers
contribute at $v k \ll \omega.$ One should note however, that $0$
in condition $0<\omega-v k\ll \omega,$ is an artifact of the
linear spectrum approximation, which breaks down at energy scales
on the order of $\varepsilon_1(k)-\varepsilon_2(k)\sim k^2.$ It is
precisely within this region where exact results
(\ref{nonluttexp2}),(\ref{nonluttexp3}) are applicable. Each of
these exponents is valid in the vicinities of
$\varepsilon_{1(2)}(k)$ much narrower than
$\varepsilon_1(k)-\varepsilon_2(k)\sim k^2.$ These exponents
describe the response of the system at $k\rightarrow 0$ beyond
linear hydrodynamic approximation. Nevertheless, for the
Lieb-Liniger model  they turn out to depend only on the LL
parameter $K,$ since phase shifts at quasifermi surface are also
related to $K;$ see Eq. (\ref{F1exact}). Far away from
$k\rightarrow 2 \pi D n ,$ regions of validity of
exponents~(\ref{muup}),(\ref{mudown}) widen, and the
singularities constitute the main features of the spectral
function.

To summarize, we have  found  exact exponents characterizing the
singularities of the dynamic structure factor and spectral
function along the dispersion curves of the collective modes of a
1D Bose gas. The found exponents are related to the known Bethe
ansatz solution of the Lieb-Liniger model. Remarkably, the
Luttinger liquid theory  predictions for the exponents fail even
at low energies, once the immediate vicinities of the edges are
considered.

 We thank  A. Kamenev for
 useful
 discussions. This work was
supported by NSF grants No. DMR-0749220 and No. DMR-0754613.

{\it Note added.} - After the essential part of this work has been
completed, preprint to Ref. \cite{CheianovPustilnik} has appeared,
where equations similar to our Eqs. (\ref{FBeq}) and (\ref{mu12})
were also reported for DSF of 1D fermions on a lattice.


\begin{thebibliography}{10}



\bibitem{dweiss} T. Kinoshita, T. Wenger, and D. S. Weiss, Nature  {\bf 440}, 900 (2006).
\bibitem{LL}E.H. Lieb and W. Liniger, Phys. Rev. {\bf 130}, 1605
(1963); E.H. Lieb, {\it ibid}. {\bf 130}, 1616 (1963).





\bibitem{KBI}V.E. Korepin, N.M. Bogoliubov, and A.G. Izergin, {\it Quantum Inverse Scattering Method and Correlation Functions}
 (Cambridge University Press, Cambridge, 1993).



\bibitem{CauxDSF} J.-S. Caux and  P. Calabrese, Phys. Rev. A {\bf 74, } 031605(R)
(2006).
\bibitem{CauxG}J.-S. Caux, P. Calabrese, and  N.A. Slavnov, J. Stat. Mech. (2007)
P01008. 
This work considered 
$A(k,\omega<0).$

\bibitem{Pustilnik2006Fermions}M. Pustilnik, M. Khodas, A. Kamenev, and L.I. Glazman, Phys. Rev. Lett. {\bf 96}, 196405 (2006).
\bibitem{Khodas2006Fermions}M. Khodas, M. Pustilnik, A. Kamenev, and L.I. Glazman, Phys. Rev. B {\bf 76}, 155402
(2007).
\bibitem{Khodas2007Bosons}M. Khodas, M. Pustilnik, A. Kamenev, and L.I. Glazman, Phys. Rev. Lett. {\bf  99}, 110405
(2007).
\bibitem{Khodas2007Soliton}M. Khodas,  A. Kamenev, and L.I. Glazman, arXiv:0710.2910v1.

\bibitem{Affleck2007} R.G. Pereira, S.R. White, and I. Affleck,
Phys. Rev. Lett. {\bf 100}, 027206 (2008).



\bibitem{EL} K.B. Efetov and A.I. Larkin, Sov. Phys. JETP {\bf 42}, 390 (1975)
.
\bibitem{Haldane} F.D.M. Haldane, Phys. Rev. Lett. {\bf 47}, 1840 (1981).
\bibitem{Caza04}M.A. Cazalilla, J. Phys. B  {\bf 37}, S1 (2004).
\bibitem{Giamarchibook}T. Giamarchi, {\it Quantum Physics in One Dimension}
(Clarendon, Oxford, 2004).
\bibitem{DMRG} U. Schollw\"{o}ck, Rev. Mod. Phys.  {\bf 77}, 259 (2005).


\bibitem{MITbraggscattering} D.M. Stamper-Kurn {\it et al.},
Phys. Rev. Lett. {\bf 83}, 2876 (1999).
\bibitem{Duan}L.-M. Duan, Phys. Rev. Lett. {\bf 96}, 103201 (2006).
\bibitem{Carusotto} T.-L. Dao, A. Georges, J. Dalibard, C. Salomon, and I.~Carusotto, Phys. Rev. Lett. {\bf 98}, 240402 (2007).
\bibitem{AGD} A.A. Abrikosov, L.P. Gorkov, and I.E. Dzyaloshinski,
 {\it Methods of Quantum Field Theory in Statistical Physics} (Dover, New York, 1963).
\bibitem{fnote} Shift function $F(\nu|\lambda)$ defined by Eq. (I.4.25) of Ref. \cite{KBI} is related to ours
as $F(\nu|\lambda)= F_B(\nu|\lambda) - \pi \rho(\nu).$ Ref.
\cite{KBI} assumes a change of periodic boundary conditions to
antiperiodic ones with the change of particle number (see p.23),
and phase shifts calculated there correspond to the fermionic
Cheon-Shigehara model \cite{Cheon1999} dual to the bosonic
Lieb-Liniger model.
\bibitem{Cheon1999}T. Cheon and T. Shigehara, Phys. Rev. Lett. {\bf 82}, 2536
 (1999).
\bibitem{KorepinSlavnov}V. Korepin and N. Slavnov, Eur. Phys. J. B  {\bf 5}, 555 (1998).
\bibitem{Schotte}K.D. Schotte and U. Schotte, Phys. Rev. {\bf 182}, 479
(1969).
\bibitem{Balents2000} L. Balents, Phys. Rev. B {\bf 61}, 4429
(2000).
\bibitem{GNT}A. Gogolin, A. Nersesyan, and A. Tsvelik, { \it Bosonization and Strongly Correlated Systems}  (Cambridge University Press, Cambridge,  1998).
\bibitem{CheianovPustilnik} V.V. Cheianov and  M. Pustilnik,
Phys. Rev. Lett. {\bf 100}, 126403 (2008).




\end{thebibliography}
\end{document}